\documentclass[aps, pre, amsmath, reprint]{revtex4-1}
\usepackage[utf8]{inputenc}
\usepackage[T1]{fontenc}
\usepackage{siunitx}
\usepackage{bm}

\newcommand{\vect}[1]{\bm{#1}}

\newcommand{\emath}{\,\mathrm{e}}
\renewcommand{\imath}{\mathrm{i}}

\newcommand{\diff}[1]{\mathop{\mathrm{d}\!}#1} 
\newcommand{\pderiv}[3][]{\frac{\partial^{#1}#2}{\partial #3^{#1}}} 

\begin{document}

\title[SCHME]{Acoustic multipole sources from the Boltzmann equation}
\author{Erlend Magnus Viggen}
\email{erlend.viggen@ntnu.no}
\affiliation{Acoustics Research Center, Department of Electronics and Telecommunications, NTNU, 7034 Trondheim, Norway}
\date{\today}

\begin{abstract}
By adding a particle source term in the Boltzmann equation of kinetic theory, it is possible to represent particles appearing and disappearing throughout the fluid with a specified distribution of particle velocities. By deriving the wave equation from this modified Boltzmann equation via the conservation equations of fluid mechanics, multipole source terms in the wave equation are found. These multipole source terms are given by the particle source term in the Boltzmann equation. To the Euler level in the momentum equation, a monopole and a dipole source term appear in the wave equation. To the Navier-Stokes level, a quadrupole term with negligible magnitude also appears.
\end{abstract}

\maketitle

\section{Introduction}

Acoustic multipoles are oscillating sources that emit acoustic fields of different directivities. These sources can be either point sources, localized at single points in space, or they can be distributions throughout the medium. The the first three orders of multipoles are the most well-known: \emph{Monopoles}, \emph{dipoles}, and \emph{quadrupoles} at zeroth, first, and second order, respectively.

When these three types of multipole sources appear as source terms in the wave equation, they usually originate from terms in the conservation equations of fluid mechanics. For instance, monopoles are linked to a source term in the mass conservation equation (also known as the continuity equation), which represents mass appearing and disappearing throughout the fluid as a function of time. This can model pulsations of small bodies throughout the fluid~\cite{howe03}.

However, an alternative approach is to add a particle source term in the \emph{Boltzmann equation}, which is more fundamental than the fluid conservation equations which can be derived from it. Adding such a source term to the Boltzmann equation allows specifying the velocity distribution of particles that appear and disappear throughout the fluid. This approach is therefore similar to, but more general than, the aforementioned method of adding a mass source term, and can therefore model more general vibrations of small bodies in the fluid.

Such a particle source term was recently examined for the lattice Boltzmann method~\cite{viggen13}, a computational fluid dynamics method based on the fully discretised Boltzmann equation. It was found that such an approach results in a wave equation with non-vanishing monopole, dipole, and quadrupole source terms. This article will similarly examine particle source terms, but in the classic non-discretised Boltzmann equation.

\section{Acoustic multipole sources}

Mathematically, multipoles are related to source terms in the wave equation,
\begin{equation} \begin{split} \label{eq:wavesourceterms}
	\left( \frac{1}{c_0^2} \pderiv{}{t} - \nabla^2 \right) p(\vect{x},t)
	&= T_0(\vect{x},t) +	\pderiv{}{x_i} T_i(\vect{x},t) \\
	&\quad + \frac{\partial^2}{\partial x_i \partial x_j} T_{ij}(\vect{x},t) + \ldots \; .
\end{split} \end{equation}
Here, $p$ is the pressure and $c_0$ is the speed of sound. The terms on the right-hand side are multipole source terms.

This article makes use of the \emph{index notation} commonly used in the field of fluid mechanics. In this notation, a single index indicates a generic vector element (e.g.\ $T_i$ could be $T_x$, $T_y$, or $T_z$), and multiple indices (as in $T_{ij}$ or $a_i b_j$) indicate generic tensor elements. Repeating indices within a single term implies summation over all possible values of that index. For example, $a_i b_i = a_x b_x + a_y b_y + a_z b_z = \vect{a}\cdot\vect{b}$, and $\partial T_i / \partial x_i = \partial T_x / \partial x + \partial T_y / \partial y + \partial T_z / \partial z = \nabla \cdot \vect{T}$. When indices repeat in this way, the letter used is arbitrary, so that $a_i b_i = a_k b_k$.

The general three-dimensional solution to~\eqref{eq:wavesourceterms} is given by an integral over the entire volume of the source terms on the right-hand side~\cite{howe03},
\begin{equation} \begin{split} \label{eq:wavesourcesolution}
	p(\vect{x},t) = \frac{1}{4\pi} \int \Bigg[ & \frac{ T_0(\vect{y},t-\tfrac{|\vect{x}-\vect{y}|}{c_0}) }{|\vect{x}-\vect{y}|}\\
	&+ \frac{\partial}{\partial x_i} \frac{ T_i(\vect{y},t-\tfrac{|\vect{x}-\vect{y}|}{c_0}) }{|\vect{x}-\vect{y}|} \\
	&+ \frac{\partial^2}{\partial x_i \partial x_j} \frac{ T_{ij}(\vect{y},t-\tfrac{|\vect{x}-\vect{y}|}{c_0}) }{|\vect{x}-\vect{y}|}
	\Bigg] \diff{\vect{y}} .
\end{split} \end{equation}
Thus, $T_0(\vect{x},t)$ indicates the instantaneous monopole strength, $T_i(\vect{x},t)$ the $i$-dipole strength, and $T_{ij}$ the $ij$-quadrupole strength.

As mentioned above, monopole sources can be modeled by adding a mass source term to the mass conservation equation,
\begin{equation} \label{eq:massconservation}
	\frac{\partial \rho}{\partial t} + \nabla \cdot \left( \rho \vect{u} \right) = Q,
\end{equation}
$\rho$ being the mass density, $\vect{u}$ the fluid velocity, and $Q(\vect{x},t)$ the instantaneous mass flux. Dipoles typically originate from the force term in the momentum conservation equation. To the Euler level, this equation is
\begin{equation} \label{eq:eulermomentum}
	\rho \left[ \frac{\partial \vect{u}}{\partial t} + (\vect{u} \cdot \nabla) \vect{u} \right] = - \nabla p + \vect{F},
\end{equation}
where $\vect{F}$ represents body forces. Finally, quadrupoles typically originate from the nonlinear term in~\eqref{eq:eulermomentum}.

The multipole terms in~\eqref{eq:wavesourceterms} are usually related to the terms in the conservation equations as~\cite{howe03}
\begin{equation*}
	T_0 = \frac{\partial Q}{\partial t}, \qquad
	T_i = - F_i, \qquad
	T_{ij} \simeq \rho u_i u_j .
\end{equation*}

\section{The Boltzmann equation}

The Boltzmann equation describes motion of a gas at a finer level of detail than the fluid conservation equations. In this discussion we shall restrict ourselves to its very basics. More details can be found in the literature~\cite{cercignani88, chapman70, hanel04}. The equation evolves the particle \emph{distribution function}, $f(\vect{x},\vect{\xi},t)$, which may be seen as a double density in both physical space and particle velocity space. Thus, it describes the density of particles with position $\vect{x}$ and velocity $\vect{\xi}$ at time $t$.

The familiar macroscopic quantities can be recovered as \emph{moments} of $f$, i.e.\ by weighting with some function and integrating over the entire velocity space. The mass density and momentum density are found as the zeroth- and first-order moments,
\begin{subequations} \label{eq:moments}
\begin{align}
	\rho(\vect{x},t) &= \int f(\vect{x},\vect{\xi},t) \diff{\vect{\xi}} , \\
	\rho \vect{u}(\vect{x},t) &= \int \vect{\xi} f(\vect{x},\vect{\xi},t) \diff{\vect{\xi}} .
\end{align}
\end{subequations}

Neglecting body forces and using the BGK collision operator~\cite{bhatnagar54}, the Boltzmann equation is
\begin{equation} \label{eq:boltzmann}
	\pderiv{f}{t} + \vect{\xi} \cdot \nabla f = s - \frac{1}{\tau} \left( f - f^{(0)} \right) .
\end{equation}
$s(\vect{x},\vect{\xi},t)$ is the aforementioned particle source term which is central to this article. As the left-hand side of the equation is a standard advection equation, $s(\vect{x},\vect{\xi},t)$ describes the rate at which particles are added into the $f(\vect{x},\vect{\xi},t)$ distribution. The final term is the BGK collision operator, which models collisions between particles as a relaxation with a characteristic \emph{relaxation time} $\tau$ to the equilibrium distribution function,
\begin{equation} \label{eq:eqdist}
	f^{(0)}(\vect{x},\vect{\xi},t) = \rho \left( \frac{\rho}{2 \pi p} \right)^{3/2} \emath^{- \rho|\vect{\xi}-\vect{u}|^2 / 2 p} ,
\end{equation}
with $\rho$ and $\vect{u}$ found from~\eqref{eq:moments}. In this article, $p$ is approximated by~\eqref{eq:isentropicpressure}. As the quantities of mass and momentum are conserved in collisions, substituting $f^{(0)}$ for $f$ in~\eqref{eq:moments} must give the same moments. As a result of this, the BGK collision operator conserves both mass and momentum.

The BGK collision operator is a far simpler model of collisions than Boltzmann's original and more accurate collision operator. That simplicity comes with drawbacks, chiefly that the BGK operator slightly mispredicts the Prandtl number~\cite{cercignani88}. This dimensionless number relates the transport coefficients in the momentum equation (viscosity) and the energy equation (conductivity). However, this will not matter as we will neglect the effects of conductivity in this article.

From this point on we will assume that the macroscopic variables fluctuate only slightly around rest state values $\rho=\rho_0$, $p=p_0$, $\vect{u}=0$. This allows us to linearise subsequent equations in this article, which is in keeping with the usual assumptions in acoustics.

The pressure $p$ in~\eqref{eq:eqdist} is approximated using the common isentropic relation~\cite{howe03, thompson72}
\begin{equation} \label{eq:isentropicpressure}
	\frac{p}{p_0} = \left( \frac{\rho}{\rho_0} \right)^\gamma ,
\end{equation}
$\gamma = (d+2)/2$ being the adiabatic index determined by the degrees of freedom $d$ of the molecules that make up the gas~\cite{thompson72}. In this way, we include the effect of equipartition of energy between translational and inner (i.e.\ rotational and vibrational) degrees of freedom, in the limit of rapid energy transfer. This relation leads to an ideal speed of sound $c_0$ given by
\begin{equation} \label{eq:soundspeed}
	c_0^2 = \pderiv{p}{\rho} = \frac{\gamma p_0}{\rho_0} .
\end{equation}

It will be useful to introduce an abbreviated notation for the moments of the particle source term $s$,
\begin{subequations} \label{eq:sourcemoments}
\begin{align}
	S_0(\vect{x},t) &= \int s(\vect{x},\vect{\xi},t) \diff{\vect{\xi}} , \\
	S_i(\vect{x},t) &= \int \xi_i s(\vect{x},\vect{\xi},t) \diff{\vect{\xi}} , \\
	S_{ij}(\vect{x},t) &= \int \xi_i \xi_j s(\vect{x},\vect{\xi},t) \diff{\vect{\xi}} ,
\end{align}
\end{subequations}
and so forth. $S_0(\vect{x},t)$ represents the instantaneous mass flux of the particles at $\vect{x}$, $S_i$ is associated with odd symmetries of $s$ in velocity space, and $S_{ij}$ is similarly associated with various even symmetries.

\section{Fluid conservation equations}

It is possible to find the conservation equations of fluid mechanics from the Boltzmann equation~\eqref{eq:boltzmann}. To find the Euler equations, we could simply take the zeroth and first moments of~\eqref{eq:boltzmann} under the assumption that $f \simeq f^{(0)}$. However, to find the momentum equation to the Navier-Stokes level, so that it includes the stress tensor term, we must resort to the \emph{Chapman-Enskog expansion}. This is a technique used to derive the fluid conservation equations from the Boltzmann equation. It is discussed throughout the literature with varying approaches and varying levels of complexity~\cite{cercignani88, chapman70, grad63, hanel04, dellar01}. In the following derivation, we use a moment-based approach~\cite{dellar01}.

Two mathematical techniques are used in this expansion. First, the distribution function $f$ is approximated as a perturbation expansion around equilibrium $f^{(0)}$ in a smallness parameter $\epsilon$. Second, a multi-scale expansion of time is performed. In mathematical notation,
\begin{align*}
	f &= f^{(0)} + \epsilon f^{(1)} + \epsilon^2 f^{(2)} + \ldots \; , \\
	\pderiv{}{t} &= \pderiv{}{t_0} + \epsilon \pderiv{}{t_1} + \ldots \; .
\end{align*}
The smallness parameter $\epsilon$ is associated with the dimensionless Knudsen number $\mathrm{Kn} = l_\mathrm{mfp} / L$, relating the mean free path $l_\mathrm{mfp}$ in the gas to a macroscopic length scale $L$. Thus, $f^{(n+1)}$ is of one order higher in the Knudsen number than $f^{(n)}$. A dimensional analysis~\cite{hanel04} reveals that the relaxation time is also at first order of smallness, so that $\tau = \epsilon \tau$. As previously explained, the density and momentum is fully contained in $f^{(0)}$, so that
\begin{equation}
	\int f^{(n)} = \int \xi_i f^{(n)} = 0 \qquad \text{for } n > 0 .
\end{equation}

Expanding the Boltzmann equation in this way and truncating the expansion to $\mathcal{O}(\epsilon)$, we find
\begin{equation} \begin{split}
	\left( \pderiv{}{t_0} + \epsilon \pderiv{}{t_1} + \xi_i \pderiv{}{x_i} \right) \left( f^{(0)} + \epsilon f^{(1)} \right) \\
	= s - \frac{1}{\epsilon\tau} \left( \epsilon f^{(1)} + \epsilon^2 f^{(2)} \right) .
\end{split} \end{equation}
Gathering these terms according to their order of smallness, we find
\begin{subequations} \label{eq:smallnessterms}
\begin{align}
	\mathcal{O}(\epsilon^0)\!: && \left( \pderiv{}{t_0} + \xi_i \pderiv{}{x_i} \right) f^{(0)} &= s - \frac{1}{\tau} f^{(1)} , \label{eq:smallness0terms} \\
	\mathcal{O}(\epsilon^1)\!: && \!\!\!\!\! \pderiv{f^{(0)}}{t_1} + \left( \pderiv{}{t_0} + \xi_i \pderiv{}{x_i} \right) f^{(1)} &= - \frac{1}{\tau} f^{(2)} . \label{eq:smallness1terms}
\end{align}
\end{subequations}
To derive the mass and momentum equations to the Euler level, only the $\mathcal{O}(\epsilon^0)$ terms are needed. The Navier-Stokes corrections to these equations are found by also including the $\mathcal{O}(\epsilon^1)$ terms in the derivation. Similarly, it is possible to find further corrections at $\mathcal{O}(\epsilon^2)$ (known as the Burnett corrections) and beyond (super-Burnett), although these further corrections are negligible in practical cases~\cite{chapman70, grad63} and have historically been viewed with some suspicion~\cite{grad63}.

Deriving the conservation equations from~\eqref{eq:smallnessterms} requires the second and third moments of $f^{(0)}$~\cite{dellar01}. In linearised form, these are
\begin{subequations} \label{eq:eqmoments}
\begin{align}
	\int \xi_i \xi_j f^{(0)} \diff{\vect{\xi}} &\simeq p \delta_{ij} , \\
	\int \xi_i \xi_j \xi_k f^{(0)} \diff{\vect{\xi}} &\simeq p_0 ( u_i \delta_{jk} + u_j \delta_{ik} + u_k \delta_{ij} ) ,
\end{align}
\end{subequations}
where $\delta_{ij}$ is the Kronecker delta.

Taking the zeroth to second moments of~\eqref{eq:smallness0terms}, using~\eqref{eq:eqmoments}, and linearising, we find
\begin{subequations}
\begin{align}
	\pderiv{\rho}{t_0} + \rho_0 \pderiv{u_i}{x_i} &= S_0 , \label{eq:small0mom0} \\
	\rho_0 \pderiv{u_i}{t_0} + \pderiv{p}{x_i} &= S_i , \label{eq:small0mom1} \\
	\delta_{ij} \pderiv{p}{t_0} + p_0 \left( \pderiv{u_i}{x_j} + \pderiv{u_j}{x_i} + \delta_{ij} \pderiv{u_k}{x_k} \right) &= S_{ij} - \frac{1}{\tau} \Pi_{ij}^{(1)} , \label{eq:small0mom2}
\end{align}
\end{subequations}
where $\Pi_{ij}^{(1)} = \int \xi_i \xi_j f^{(1)} \diff{\vect{\xi}}$. With the $\mathcal{O}(\epsilon^0)$ approximation $\partial / \partial t_0 = \partial / \partial t$, the two first equations are equivalent to linearised versions of the mass equation~\eqref{eq:massconservation} and the Euler momentum equation~\eqref{eq:eulermomentum}, with $S_0$ and $S_i$ in the place of $Q$ and $F_i$, respectively. Taking the zeroth and first moments of~\eqref{eq:smallness0terms} and linearising, we find $\mathcal{O}(\epsilon)$ corrections to the above equations,
\begin{subequations}
\begin{align}
	\pderiv{\rho}{t_1} &= 0 , \label{eq:small1mom0} \\
	\rho_0 \pderiv{u_i}{t_1} + \pderiv{\Pi_{ij}^{(1)}}{x_j} &= 0 . \label{eq:small1mom1}
\end{align}
\end{subequations}

The sum $\eqref{eq:small0mom0} + \epsilon \eqref{eq:small1mom0}$ directly leads to the linearised mass equation, given below in~\eqref{eq:massconservation2}. Similarly, $\eqref{eq:small0mom1} + \epsilon \eqref{eq:small1mom1}$ leads to the momentum equation, though the unknown tensor $\Pi^{(1)}_{ij}$.

$\Pi^{(1)}_{ij}$ can be explicitly related to the Navier-Stokes stress tensor using~\eqref{eq:small0mom2}. The pressure time derivative can be rewritten assuming a nearly isentropic process and using~\eqref{eq:small0mom0}, resulting in
\begin{equation} \label{eq:densitytopressure}
	\pderiv{p}{t_0} = \pderiv{p}{\rho} \pderiv{\rho}{t_0} = c_0^2 \left( S_0 - \rho_0 \pderiv{u_k}{x_k} \right) .
\end{equation}
Substituting for the speed of sound using~\eqref{eq:soundspeed}, the diagonal terms in~\eqref{eq:small0mom2} become
\begin{equation} \begin{split}
	\pderiv{p}{t_0} + p_0 \pderiv{u_k}{x_k}
	&= \left( p_0 - \rho_0 c_0^2 \right) \pderiv{u_k}{x_k} + c_0^2 S_0 \\
	&= - p_0 ( \gamma - 1 ) \pderiv{u_k}{x_k} + c_0^2 S_0 .
\end{split} \end{equation}
Thus, we find
\begin{equation} \begin{split}
	\Pi_{ij}^{(1)} 
	&= - p_0 \tau \left[ \pderiv{u_j}{x_i} + \pderiv{u_i}{x_j} + \delta_{ij} (1-\gamma) \pderiv{u_k}{x_k} \right] \\
	&\quad + \tau \left( S_{ij} - c_0^2 S_0 \right) .
\end{split} \end{equation}

The mass and momentum equations can now be explicitly found as described above,
\begin{subequations} \label{eq:conservation2}
\begin{align}
	\pderiv{\rho}{t} + \rho_0 \pderiv{u_i}{x_i} &= S_0 , \label{eq:massconservation2} \\
	\rho_0 \pderiv{u_i}{t} + \pderiv{p}{x_i}
	&= S_i - \frac{\mu}{p_0} \pderiv{(S_{ij} - \delta_{ij} c_0^2 S_0)}{x_j} + \pderiv{\sigma'_{ij}}{x_j} . \label{eq:momentumconservation2}
\end{align}
\end{subequations}
The momentum equation contains a deviatoric stress tensor
\begin{equation}
	\sigma'_{ij} = \mu \left( \pderiv{u_i}{x_j} + \pderiv{u_j}{x_i} - \frac{2}{3} \delta_{ij} \pderiv{u_k}{x_k} \right) + \mu_B \delta_{ij} \pderiv{u_k}{x_k} ,
\end{equation}
with shear viscosity $\mu = p_0 \tau$ and bulk viscosity $\mu_B/\mu = (5/3 - \gamma)$. This value of the bulk viscosity in the limit of rapid transfer of energy between translational and inner degrees of freedom has previously been found using more rigorous kinetic theory~\cite{morse64}.

Comparing~\eqref{eq:massconservation2} to the classical mass equation~\eqref{eq:massconservation}, $S_0$ appears in the place of the mass flux $Q$, which could be expected considering the interpretation of $S_0$ as a mass flux. Comparing~\eqref{eq:momentumconservation2} to the Euler-level momentum equation~\eqref{eq:eulermomentum}, $S_i$ appears in the place of the body force, which had been neglected from the Boltzmann equation. $S_{ij}$ is also present inside a source term with a single spatial derivative and a small coefficient $\mu/p_0$ in front.

\section{The wave equation}

The wave equation can be found as usual from the mass and momentum equations as $\partial\eqref{eq:massconservation2} / \partial t - \partial\eqref{eq:momentumconservation2} / \partial x_i$. Using the isentropic relation~\eqref{eq:densitytopressure} in the wave equation operator, and thus neglecting sound absorption due to thermal conduction and relaxation~\cite{blackstock00}, we find
\begin{align}
	\left( \frac{1}{c_0^2} \pderiv[2]{}{t} - \nabla^2 \right) p
	&= \pderiv{S_0}{t} - \pderiv{S_i}{x_i} 
	+ \frac{\mu}{p_0} \frac{\partial^2 (S_{ij} - \delta_{ij} c_0^2 S_0)}{\partial x_i \partial x_j} \nonumber \\
	&\quad+ \frac{\partial^2 \sigma'_{ij}}{\partial x_i \partial x_j} .
\end{align}
Comparing with~\eqref{eq:wavesourceterms} and~\eqref{eq:wavesourcesolution}, we find a monopole strength $T_0 = \partial S_0 / \partial t$, a dipole strength $T_i = - S_i$, and a quadrupole strength which is not fully resolved but involves $S_{ij}$, $\delta_{ij} S_0$, and $\sigma'_{ij}$.

The deviatoric stress tensor $\sigma'_{ij}$ can be resolved as
\begin{equation}
	\frac{\partial^2 \sigma'_{ij}}{\partial x_i \partial x_j}
	= \mu (3 - \gamma) \nabla^2 \pderiv{u_k}{x_k}
	= \frac{\mu (3 - \gamma)}{\rho_0} \nabla^2 \left( S_0 - \frac{1}{c_0^2} \pderiv{p}{t} \right) .
\end{equation}
The last term in the parenthesis contributes to sound absorption and is neglected in line with previous approximations. The first parenthetical term contributes to the quadrupole strength. Using the property $\nabla^2 = \delta_{ij} (\partial^2 / \partial x_i \partial x_j)$, we find a fully resolved isentropic wave equation
\begin{equation} \begin{split}
	\left( \frac{1}{c_0^2} \pderiv[2]{}{t} - \nabla^2 \right) p
	&= \pderiv{S_0}{t} - \pderiv{S_i}{x_i} \\
	&\quad + \frac{\mu}{p_0} \frac{\partial^2 (S_{ij} - 3 \delta_{ij} p_0 S_0 / \rho_0)}{\partial x_i \partial x_j} .
\end{split} \end{equation}
Finally, we find the quadrupole strength as $T_{ij} = (\mu / p_0) (S_{ij} - 3 \delta_{ij} p_0 S_0 / \rho_0)$.

The coefficient $\mu/p_0$ in front of the quadrupole strength is typically on the order of $\SI{E-10}{\second}$ in gases. Its small magnitude means that the quadrupoles generated by the Boltzmann equation source term $s$ tend to be negligible compared to the monopoles and dipoles.

\section{Summary and conclusion}

A source term $s$ in the Boltzmann equation represents particles appearing or disappearing throughout the fluid with some distribution of particle velocities. As adding a source term to the mass equation allows modeling pulsations of small bodies throughout the fluid~\cite{howe03}, a source term in the Boltzmann equation would allow modeling more general vibrations of such small bodies.

From this modified Boltzmann equation, the mass and momentum conservations equations~\eqref{eq:conservation2} were derived under the common acoustic approximation of linearity and constant entropy. These equations gain source terms given by the moments~\eqref{eq:sourcemoments} of $s$. The mass equation gains a source term $S_0$. To the Euler level, the momentum equation gains a source term $S_i$, and to the Navier-Stokes level, it gains a source term involving $S_{ij}$ and $\delta_{ij} S_0$.

The wave equation derived from these two conservation equations contains multipole source terms. The monopole strength is $\partial S_0 / \partial t$ and the dipole strength is $-S_i$. The quadrupole strength, which comes out of the Navier-Stokes level of the momentum equation, involves $S_{ij}$ and $\delta_{ij} S_0$, but has a negligible magnitude. That the quadrupole strength is so much smaller than the monopole and dipole strength could be expected, as the Navier-Stokes level terms are one order higher in the small Knudsen number than the Euler-level terms.

Similarly, going to the $\mathcal{O}(\mathrm{Kn}^2)$ Burnett level might lead to a $\partial^2 S_{ijk} / \partial x_j \partial x_k$ term in the momentum equation, leading to an octupole term in the wave equation. However, since this term would be at $\mathcal{O}(\mathrm{Kn}^2)$, it would be even more negligible than the quadrupole term.

Going back to the point of modeling general vibrations of small bodies throughout the fluid, this analysis indicates that the vibrations of such small bodies can radiate as monopoles and dipoles, but only very weakly as higher-order multipoles.

Comparing this analysis with the analogous analysis for the lattice Boltzmann method~\cite{viggen13}, we find that the analogous source term in the lattice Boltzmann equation can only radiate quadrupoles effectively due to a fortuitous discretisation error that occurs when discretising the Boltzmann equation~\eqref{eq:boltzmann} in space and time using the first-order rectangle method. Discretising with the trapezoidal method~\cite{dellar01}, which results in a scheme fully consistent with~\eqref{eq:boltzmann}, would therefore also lead to a vanishing quadrupole term similarly to what has been shown here.

\bibliographystyle{apsrev4-1}
\bibliography{boltzmannmultipoles}

\end{document}